\documentclass[]{article}

\title{Merger of vorticity holes}

\author{Oscar Velasco Fuentes\\
            Departamento de Oceanograf\'{\i}a F\'{\i}sica\\
            CICESE, Ensenada, M\'exico,
            ovelasco@cicese.mx      
}
\date{13 October 2010}

\begin{document}

\maketitle

\begin{abstract}
This paper contains background information for a fluid dynamics video
submitted to the Gallery of Fluid Motion to be held along with the  63rd Annual Meeting 
of the American Physical Society's Division of Fluid Dynamics.
\end{abstract}

A vorticity hole is a compact region that has vorticity which is, in absolute value, significantly lower
than that of its surroundings. 
They have been observed under the most varied conditions, such as electron plasmas 
(e.g. Durkin and Fajans 2000),  and in the Earth's atmosphere  (e.g. Knaff et al. 2008).

This video shows the evolution, computed numerically with a contour dynamics model,
of two identical vorticity holes within a circular vortex. 
The vorticity is positive and uniform, say $\Omega$, within the yellow region; 
it decreases in equal steps ($\Omega/4$) as the color becomes darker and it is zero in the black regions. 

In the initial condition the hole radius is one fifth of the vortex radius,
and the holes are separated by a distance equal to three times their own radius.
This configuration is not steady, therefore the two vorticity 
holes rapidly approach each other and undergo strong deformations. 
Then they merge into a single, elongated hole while forming two vorticity-hole filaments.
Although very similar to the merger of vortices (e.g. Velasco Fuentes 2005),
the merger of vorticity holes additionally involves the interaction with the
vortex' outer boundary. This, for instance, leads to a smaller critical distance
for merger.

\section*{References}

\parindent 0cm
\parskip 0.2cm

Durkin, D. and Fajans, J. 2000. Experimental dynamics of a vortex within a vortex.
{\em Physical Review Letters} {\bf 85}, 4052--4055.

Knaff, J.A. et al. 2008. Objective identification of annular hurricanes.
{\em Weather Forecasting} {\bf 23}, 17--28.

Velasco Fuentes, O.U. 2005. Vortex filamentation: its onset and its role on axisymmetrization and merger. 
{\em Dynamics of Atmospheres and Oceans} {\em 40}, 23-42. 

\end{document}